\documentclass[12pt]{iopart}
\usepackage{iopams,epsfig}  
\begin{document}

\title{K$^{\boldsymbol{+}}$- and K$^{\boldsymbol{-}}$-production 
       in heavy-ion collisions at SIS-energies}

\author{A. F{\"o}rster for the KaoS-Collaboration:\\
        I. B{\"o}ttcher$^{d}$, A. F{\"o}rster$^{a}$, E. Grosse$^{f,g}$, 
        P. Koczo{\'n}$^{b}$, \\
        B. Kohlmeyer$^{d}$, S. Lang$^{a}$, F. Laue$^{b,*}$,
        M. Menzel$^{d}$, \\
        L. Naumann$^{f}$, H. Oeschler$^{a}$, 
        M. P{\l}osko{\'n}$^{b}$, \\ 
        F. P{\"u}hlhofer$^{d}$, W. Scheinast$^{f}$, 
        A. Schmah$^{a}$, T. Schuck$^{c}$, \\ 
        E. Schwab$^{b}$, 
        P. Senger$^{b}$, Y. Shin$^{c}$, H. Str{\"o}bele$^{c}$, \\ 
        C.Sturm$^{a}$, 
        F. Uhlig$^{a}$, A. Wagner$^{f}$, W. Walu{\'s}$^{e}$}

\address{$^{a}$ Technische Universit{\"a}t Darmstadt,
                D-64289 Darmstadt, Germany}
\address{$^{b}$ Gesellschaft f{\"u}r Schwerionenforschung, 
                D-64291 Darmstadt, Germany} 
\address{$^{c}$ Johann Wolfgang Goethe Universit{\"a}t,
                D-60325 Frankfurt am Main, Germany}
\address{$^{d}$ Phillips Universit{\"a}t, 
                D-35037 Marburg, Germany} 
\address{$^{e}$ Uniwersytet Jagiello{\'n}ski, 
                PL-30-059 Krak{\'o}w, Poland}
\address{$^{f}$ Forschungszentrum Rossendorf,
                D-01314 Dresden, Germany}
\address{$^{g}$ Technische Universit{\"a}t Dresden,
                D-01062 Dresden, Germany}
\address{$^{*}$ Present address: Brookhaven National Laboratory, USA}

\ead{a.foerster@gsi.de}

\begin{abstract}
The production and the propagation of K$^+$- and of K$^-$-mesons 
in heavy-ion collisions at beam energies 
of 1 to 2~AGeV have systematically been investigated 
with the Kaon Spectrometer KaoS at the SIS at the GSI.
The ratio of the K$^+$-production excitation function 
for \mbox{Au+Au}  and for \mbox{C+C} reactions
increases with decreasing beam energy, which is expected 
for a soft nuclear equation-of-state.
At $1.5$~AGeV a comprehensive study of the 
K$^+$- and of the K$^-$-emission as a function of the 
size of the collision system, of the collision 
centrality, of the kaon energy, and 
of the polar emission angle has been performed.
The K$^-$/K$^+$ ratio is found to be nearly
constant as a function of the collision centrality. The spectral
slopes and the polar emission patterns are different for K$^-$ and 
for K$^+$. These observations indicate that K$^+$-mesons decouple
earlier from the reaction zone  than K$^-$-mesons.

\end{abstract}





\section{Introduction} \label{introduction}

Heavy-ion collisions provide the unique possibility to study
baryonic matter well above saturation density. The conditions
inside the dense reaction zone and the in-medium properties of
hadrons can be explored by measuring the particles created in such
collisions \cite{aich,brown}. In particular, the production of 
strange mesons at beam energies below or close to their 
respective threshold in binary nucleon-nucleon collisions 
($\rm{NN} \rightarrow \rm{K^+\Lambda N}$ at $E_{beam} = 1.6$~GeV, 
$\rm{NN} \rightarrow \rm{K^+ K^- NN}$ at  $E_{beam} = 2.5$~GeV)
is well suited for these studies.
The production at these energies  requires 
multiple nucleon-nucleon collisions or 
secondary collisions like $\rm{\pi N} \rightarrow \rm{K^+ Y}$
($\rm{Y} = \Lambda, \Sigma$) for the K$^+$ or 
the strangeness exchange reaction 
$\rm{\pi Y} \rightarrow \rm{K^- N}$ for the K$^-$.
The propagation of K$^+$-mesons in nuclear matter is 
characterized by the absence of absorption
(they contain an $\rm{\bar{s}}$-quark and hence cannot be
absorbed in hadronic matter consisting almost entirely of u- and 
d-quarks).
In combination with the sensitivity to multi-step processes, 
which are more likely to occur at high 
particle densities, this absence of absorption 
makes them nearly undisturbed 
messengers carrying information on the hot and dense 
phase of the reaction. This allows
to extract information on the nuclear equation-of-state
\cite{sturm,fuchs}.
The K$^-$ on the other hand can be reabsorbed 
via the inverse direction of the strangeness 
exchange reaction mentioned above.
By analyzing observables sensitive to the 
reaction dynamics like energy distributions and 
angular distributions, both as a function of 
the size of the reaction system and of 
the collision centrality, the time dependence 
of the K$^+$- and of the K$^-$-emission 
can be investigated.

The experiments were performed with the Kaon Spectrometer (KaoS)
at the heavy-ion synchrotron (SIS) at the GSI in Darmstadt
\cite{senger}. The magnetic spectrometer has a large acceptance in
solid angle and in momentum ($\Omega \approx 30$~msr,
$p_{max}/p_{min} \approx 2$).  The particle identification and the
trigger are based on separate measurements of the momentum and
of the time-of-flight. The trigger suppresses pions and protons by
factors of 10$^2$ and of  10$^3$, respectively. The background due to
spurious tracks and pile-up is removed by a trajectory
reconstruction based on three large-area multi-wire proportional
counters. The short distance of 5 - $6.5$~m from the target to the focal
plane minimizes the number of 
kaon-decays in flight. The loss of
kaons by decay is accounted for by Monte Carlo
simulations using the GEANT code.
Three different collision systems have been investigated
(\mbox{Au+Au}, \mbox{Ni+Ni}, and \mbox{C+C}) at 
incident beam energies ranging from $0.6$~AGeV to $2.0$~AGeV.
Results have been published in 
\cite{misko,ahner,barth,shin,laue,menzel,sturm,foerster}
and will in part be presented in this contribution.

Section \ref{eos} briefly summarizes 
the results on the excitation function of the 
K$^+$-production and the conclusions 
that can be drawn on the nuclear equation-of-state.
In section \ref{data}
a detailed analysis of dynamical observables 
for the reaction systems
\mbox{Au+Au} and \mbox{Ni+Ni} at an incident
beam energy of $1.5$~AGeV is presented, followed 
by conclusions on the time dependence 
of the emission of K$^+$- and of K$^-$-mesons 
in section \ref{conclusion}.


\section{K$^{\boldsymbol{+}}$-production as a probe 
         for the nuclear equation-of-state}  \label{eos}

Early transport calculations predicted that the K$^+$-yield in 
\mbox{Au+Au} 
collisions at beam energies below the production threshold 
in nucleon-nucleon collisions would be enhanced by a factor of about 
2 if a soft rather than a hard equation-of-state is assumed 
\cite{aich,li_ko}. Recent calculations take into account  
modifications of the kaon properties within the dense 
nuclear medium. 
The assumed repulsive K$^{+}$N-potential depends nearly (or less than)
linearly on the baryonic density \cite{schaffner} 
and thus reduces the K$^+$-yield accordingly. On the other hand, at 
subthreshold beam energies the K$^{+}$-mesons are created in secondary 
collisions involving two or more particles and hence their production  
depends at least quadratically on the density. 
To disentangle these two competing effects we have studied 
the K$^+$-production in a light (\mbox{C+C}) and in 
a heavy collision 
system (\mbox{Au+Au}) at different beam energies near threshold. 
The maximum baryonic density reached in \mbox{Au+Au}
reactions is significantly higher than in 
\mbox{C+C} reactions.
Moreover, the maximum baryonic density reached in Au+Au reactions  
depends strongly on the compression modulus of nuclear matter $\kappa$
\cite{aichelin,li_ko} whereas in C+C collisions this dependence is
rather weak \cite{fuchs}. 
Hence, the ratio of the K$^+$-multiplicity per nucleon $M/A$
in \mbox{Au+Au} to the one in \mbox{C+C} is expected 
to be sensitive to the compression modulus $\kappa$
while providing the advantage that uncertainties 
within the experimental data (beam normalization etc.) 
and within the transport model
calculations (elementary cross sections etc.) are partly cancelled.

\begin{figure}[h]
  \begin{center}
    \epsfig{file=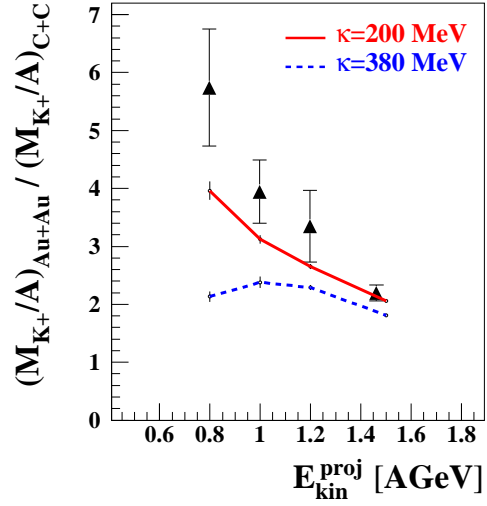,width=7.5cm}
  \end{center} 
  \caption{ The ratio of the K$^+$-multiplicitiy per nucleon 
  $M/A$ in \mbox{Au+Au} to the one in 
  \mbox{C+C} reactions as a function of the beam energy for 
  inclusive reactions. The data are compared to the results of RQMD 
  transport model calculations \cite{fuchs}. The calculations
  were performed with two different values for the compression modulus: 
  $\kappa$ = 200~MeV (a ``soft'' equation-of-state), 
  denoted by the solid line, and 
  $\kappa$ = 380~MeV (a ``hard'' EoS), indicated by the dashed line.}
  \label{fig_eos}
\end{figure}

Figure \ref{fig_eos} shows a comparison of the ratio 
$(M/A)_{\rm{Au+Au}}/(M/A)_{\rm{C+C}}$ for K$^+$ with the predictions of 
RQMD transport model calculations \cite{fuchs}. 
The calculations were performed with two values for the compression 
modulus: $\kappa=200$~MeV (solid line) and $\kappa=380$~MeV (dashed line) 
corresponding to a ``soft'' or ``hard'' nuclear equation-of-state, 
respectively. The RQMD transport model takes into account  
a repulsive K$^{+}$N-potential and uses 
momentum-dependent Skyrme forces to
determine the compressional energy per nucleon (i.e. the energy stored
in the compression) as a function of the baryon density. The comparison 
demonstrates clearly that only the calculation based on a soft nuclear 
equation-of-state reproduces the trend of the experimental data.
A similar result is obtained using the transport code IQMD \cite{har02}.

\section{K$^{\boldsymbol{+}}$- and K$^{\boldsymbol{-}}$-production
         at 1.5~AGeV}  \label{data}

In this section detailed results of experiments on 
the production and on the propagation of K$^+$- and of
K$^-$-mesons in \mbox{Ni+Ni} and in \mbox{Au+Au} collisions at a beam
energy of $1.5$~AGeV are presented. 
This is the lowest beam energy where
K$^-$ have been observed so far in collisions between heavy
nuclei. The spectral and the angular distributions of
strange mesons as a function of the collision centrality 
have been measured and significant differences 
between K$^+$ and K$^-$ have been found \cite{foerster}.

\begin{figure}[h]
  \begin{center}
    \epsfig{file=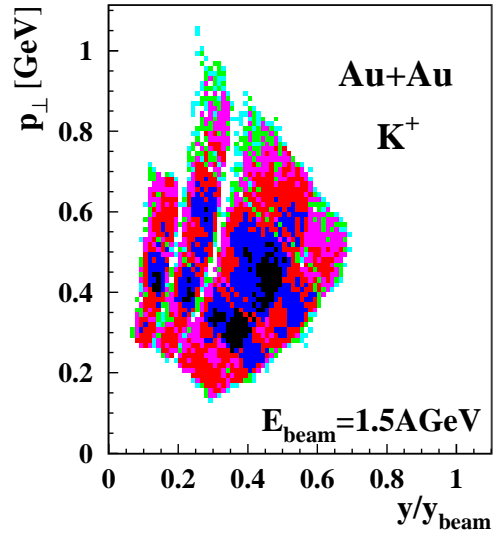,width=7.5cm}
  \end{center} 
  \caption{Transverse momentum $p_{\bot}$ 
           as a function of the rapidity $y/y_{beam}$ for K$^+$. 
           The different bands correspond 
           to different spectrometer angles 
           in the laboratory $\theta_{lab}$.}
  \label{fig_pty}
\end{figure}

Due to the energy loss in the Au-target (thickness 0.5 mm) the
average energy of the Au-beam is $1.48$~AGeV. The energy loss
of the Ni-ions in the Ni-target is negligible. A high-statistics
sample of K-mesons was recorded at a polar angle of
$\theta_{lab}$ = 40$^{\circ}$. Data from Au+Au collisions were
also taken  at $\theta_{lab}$ = 32$^{\circ}$,
48$^{\circ}$, 60$^{\circ}$, and 72$^{\circ}$ with lower statistics. 
The laboratory
momenta of the K-mesons range from 260 to 1100 MeV. 
Figure \ref{fig_pty} shows the acceptance covered 
by these measurements as a function of the rapidity 
normalized to beam rapidity $y/y_{beam}$ and of the 
transverse momentum $p_{\bot}$.

The centrality of the collision was derived from the multiplicity of
charged particles measured in the interval $12^{\circ} <
\theta_{lab} < 48^{\circ}$ by a hodoscope consisting of 84
plastic-scintillator modules.
In order to study the centrality dependence the data
measured close to midrapidity ($\theta_{lab} = 40^{\circ}$) 
were grouped into
five centrality bins both for \mbox{Ni+Ni} and for 
\mbox{Au+Au} collisions. 
The most central collisions correspond to 5\% of the total reaction
cross-section $\sigma_R$, the subsequent centrality bins
correspond to 15\%, 15\% and 25\% of $\sigma_R$. 
The most peripheral collisions correspond to 40\% of $\sigma_R$.
The total
reaction cross-section was derived from a measurement with a
minimum bias trigger and was found to be $\sigma_R$ = 6.0$\pm$0.5
barn for Au+Au and $\sigma_R$ = 2.9$\pm$0.3 barn for Ni+Ni
collisions. The corresponding mean number of participating nucleons
for each centrality bin 
$\langle A_{part} \rangle$ was calculated from the measured reaction
cross-section fraction for this bin using a geometrical model assuming
spherical nuclei.

\begin{figure}
  \begin{center}
    \mbox{\epsfig{figure=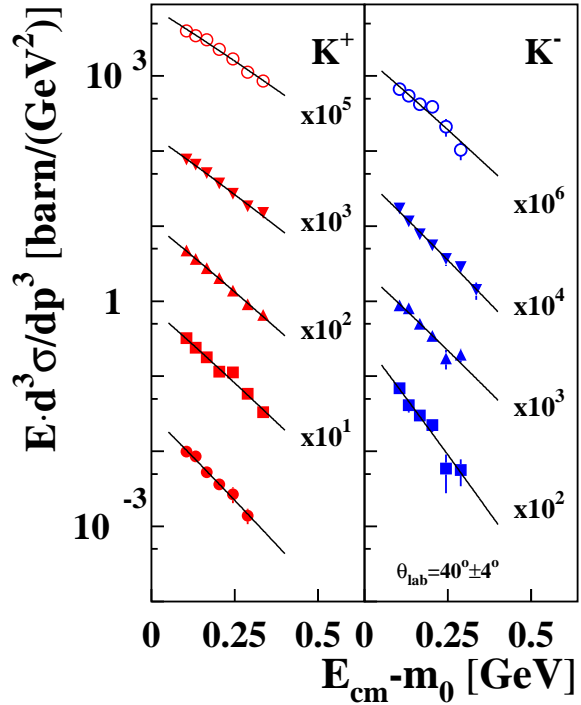,width=10cm,clip=}}
    \caption{Invariant cross sections for K$^+$ and for K$^-$ 
             in \mbox{Au+Au} collisions 
             at $E_{beam}=1.5$~AGeV for the different centrality bins. 
             The open circles depict 
             the most central data, the other bins are shown
             from the top to the bottom 
             of the figure with decreasing centrality.
             Due to low statistics the most peripheral bin 
             is not shown for K$^-$.}
    \label{fig_spec_tmbin}
  \end{center} 
\end{figure}

Figure \ref{fig_spec_tmbin} shows the production 
cross sections for K$^+$-
and for K$^-$-mesons measured close to midrapidity as a function of
the kinetic energy in the center-of-momentum system  for the five
centrality bins in Au+Au collisions. The uppermost spectra
correspond to the most central reactions, the subsequent bins 
are shown from the top to the bottom 
of the figure with decreasing centrality.  The error bars
represent the statistical uncertainties of the kaon and the
background events. An overall systematic error of 10\% due to
efficiency corrections and beam normalization has to be added. The
solid lines represent the Boltzmann function 
$d^3\sigma/dp^3 = C \cdot E \cdot \exp(-E/T)$ fitted to the data. 
$C$ is a normalization constant and the 
exponential function describes the energy
distribution with $T$ being the inverse slope parameter.

\begin{figure}
  \begin{center}
    \mbox{\epsfig{figure=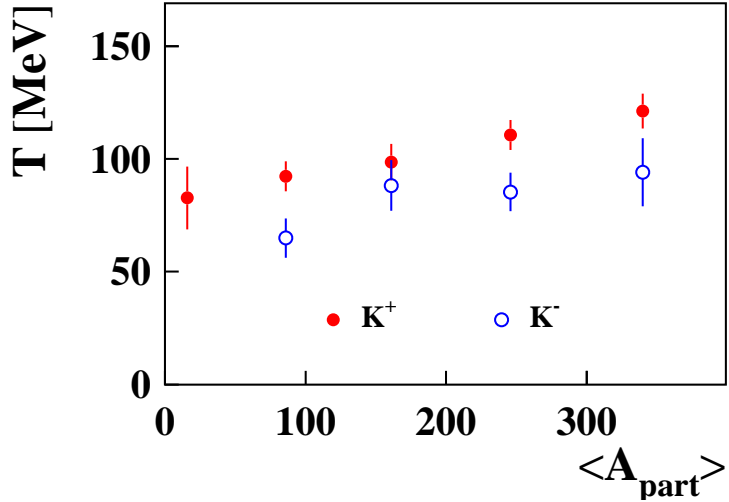,width=11cm,clip=}}
    \caption{Inverse slope parameters $T$ as a function of the 
             number of participating nucleons $\langle A_{part} \rangle$ 
             for K$^+$ (full symbols) and K$^-$ (open symbols)
             in \mbox{Au+Au} collisions at $1.5$~AGeV measured 
             at $\theta_{lab}=40^{\circ}$.}
    \label{fig_slopes}
  \end{center} 
\end{figure}

The spectra presented in figure \ref{fig_spec_tmbin} exhibit a distinct
difference between K$^-$ and K$^+$: The slopes of the 
K$^-$-spectra are steeper than those of the K$^+$-spectra. The inverse
slope parameters $T$ as a function of the mean number of participating
nucleons $\langle A_{part} \rangle$ for each centrality bin
are displayed in figure \ref{fig_slopes}. 
$T$ increases with increasing centrality and
is found to be significantly lower for K$^-$ than for K$^+$,
even for the most central collisions.

\begin{figure}[h]
  \begin{center}
  \mbox{\epsfig{figure=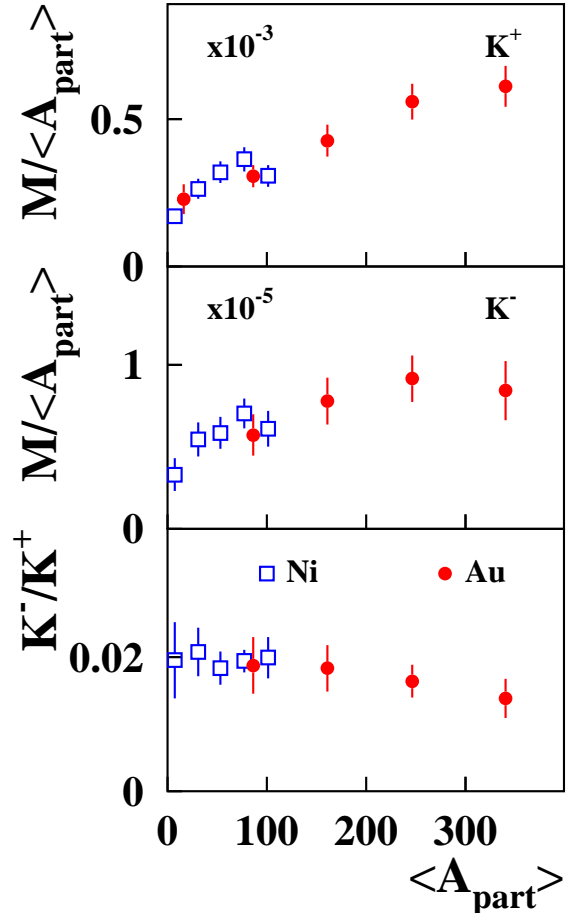,width=9cm,clip=}}
  \caption{\underline{Upper panel:} The multiplicity 
           of K$^+$ per number of participating nucleons
           $M/\langle A_{part} \rangle$ as a function of 
           $\langle A_{part} \rangle$ for \mbox{Au+Au} 
           (full circles) and \mbox{Ni+Ni} (open squares)
           at $1.5$~AGeV measured at $\theta_{lab}=40^{\circ}$.
           \underline{Middle panel:} The same but for K$^-$.
           \underline{Lower panel:} The resulting
           K$^-$/K$^+$ ratio as a function of $\langle A_{part} \rangle$.}
  \label{fig_ratio}
\end{center} 
\end{figure}

The multiplicities of K$^+$- and of K$^-$-mesons in \mbox{Ni+Ni} 
as well as in \mbox{Au+Au}
collisions at $1.5$
~AGeV differ by about a factor of 50. The
inclusive kaon multiplicity is defined for each centrality bin as
$M = \sigma_{\rm{K}}/(f\cdot\sigma_R)$ with $\sigma_{\rm{K}}$ 
being the kaon production cross
section and $(f\cdot\sigma_r)$ being the fraction of the 
reaction cross-section for the
particular event class. Figure \ref{fig_ratio} presents the 
multiplicity per number of participating nucleons 
$M/\langle A_{part}\rangle$ as a function
of $\langle A_{part} \rangle$ for K$^+$ (upper panel) 
and for K$^-$ (middle panel). 
Both, for K$^+$- and for K$^-$-mesons the multiplicities
exhibit a similar rise with $\langle A_{part} \rangle$. 
Moreover, $M/\langle A_{part} \rangle$ is
found to be almost identical in Ni+Ni and in  Au+Au collisions. 
The resulting
K$^-$/K$^+$ ratio is about 0.02 below 
$\langle A_{part} \rangle = 100$ and decreases
slightly to about 0.015 for the most central collisions
(lower panel).

\begin{figure}[h]
  \begin{center}
  \mbox{\epsfig{figure=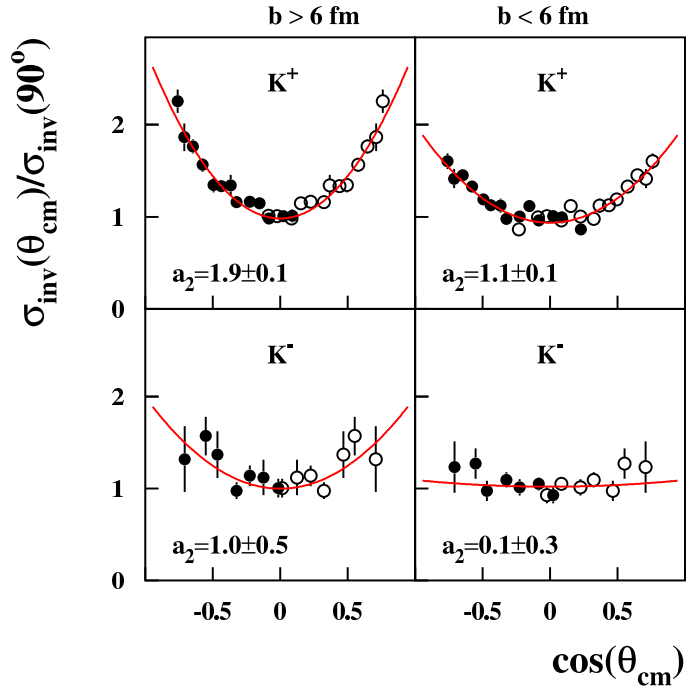,width=9cm,clip=}}
  \caption{Polar angle distributions for K$^+$ (upper panels) 
           and for K$^-$ (lower panels) in \mbox{Au+Au} 
           collisions at $E_{beam}=1.5$~AGeV. The left panels 
           show data for impact 
           parameters $b>6$~fm, the right ones for $b<6$~fm. 
           Fits and the parameter $a_{2}$ 
           are as described in the text.}
  \label{fig_angdis}
\end{center} 
\end{figure}

Another observable sensitive to the production mechanism is the
polar angle emission pattern.  The deviation from isotropy of the
K$^+$- and of the K$^-$-emission can be studied using the ratio
$\sigma_{inv}(E_{cm},\theta_{cm})$/$\sigma_{inv}(E_{cm},
90^\circ)$ as a function of $\cos(\theta_{cm})$. Here,
$\sigma_{inv}(E_{cm},\theta_{cm})$ is the invariant kaon
production cross-section measured at the polar angle $\theta_{cm}$
in the center-of-momentum frame and $\sigma_{inv}(E_{cm},
90^\circ)$ is the one measured at $\theta_{cm} = 90^\circ$. Due to
limited statistics we considered only Au+Au collisions grouped
into two centrality bins: near-central (impact parameter $b<$6 fm)
and non-central collisions ($b>$6 fm). Figure \ref{fig_angdis} displays
the anisotropy ratio for K$^+$ (upper panels) and for K$^-$ (lower
panels), both for near-central (right) and for non-central collisions
(left). For an isotropic distribution this ratio would be constant
and identical to 1.

The solid lines in figure \ref{fig_angdis} represent the function $1 +
a_2 \cdot \cos^2(\theta_{cm})$ which is fitted to the experimental
distributions with the values of $a_2$ given in the figure. In
near-central collisions the K$^-$-mesons  exhibit an isotropic
emission pattern whereas the emission of K$^+$-mesons is
forward-backward peaked. The angular distributions observed for
K$^+$ and for K$^-$ in \mbox{Ni+Ni} collisions at $1.93$~AGeV are
similar to the ones presented in  figure \ref{fig_angdis} \cite{menzel}.
The measured emission patterns indicate that the K$^-$ - in
contrast to the K$^+$ - have lost the memory of the beam direction
for central heavy-ion collisions.

\section{The time dependence of the 
          K$^{\boldsymbol{+}}$- and of the K$^{\boldsymbol{-}}$-emission}
         \label{conclusion}

When summarizing the experimental results on the 
production and on the propagation of $K$-mesons at $1.5$~AGeV
remarkable similarities
and differences between  K$^+$ and K$^-$ are found. The yields
of the K$^+$- and of the K$^-$-mesons are 
related to each other (as indicated
by the almost constant K$^-$/K$^+$ ratio as a function of
$\langle A_{part} \rangle$). 
On the other hand, the phase-space distributions (polar angles
and spectral slopes) of the K$^+$- and of the K$^-$-mesons differ
significantly. These observations can be explained by the
following scenario. In the early phase of the collision the K$^+$-mesons
and the hyperons are produced via processes like 
$\rm{NN} \rightarrow \rm{K^+ YN}$ or
$\rm{\pi N} \rightarrow \rm{K^+ Y}$ with $\rm{Y=\Lambda,\Sigma}$
\cite{fuchs,aichelin,cass_brat}. The K$^+$-mesons leave the
reaction volume with little rescattering because of their long
mean free path. Therefore, they probe the early, dense and
hot phase of the collision and have been used to obtain
information on the nuclear equation-of-state as discussed 
in section \ref{eos}.
The K$^-$-mesons, however, can hardly be produced in direct NN
collisions at a beam energy of $1.5$~AGeV, even when taking
into account Fermi motion. Indeed, transport model calculations
predict that in heavy-ion collisions at SIS energies the
production and absorption of K$^-$ predominantly proceeds via
strangeness-exchange reactions $\rm{\pi Y} \rightleftharpoons \rm{K^-N}$
\cite{Ko,cass_brat,har03}. 
As the K$^+$ are produced associated with the hyperons  
the yields of K$^+$ and K$^-$ are correlated.
Indeed, the measured K$^-$/K$^+$
ratio only slightly decreases  with increasing centrality, i.e.
with the system size.

The mean free path of the K$^-$-mesons is about $0.8$~fm in nuclear
matter due to strangeness-exchange reactions like 
$\rm{K^-N} \rightarrow \rm{Y\pi}$.
However, via the inverse reaction 
($\rm{\pi Y} \rightarrow \rm{K^- N}$) the
K$^-$ may reappear again thus propagating to the surface of
the reaction volume. Those of the K$^-$-mesons which reach the detector
are emitted in a late stage of the collision when the reaction volume is
expanding and cooling down. Consequently, the spectral slope of
the K$^-$ is steeper than the one of the K$^+$. Furthermore, the
emission pattern of the K$^-$-mesons is nearly isotropic due to
multiple collisions. Both features are observed in the experiment.

\begin{figure}[h]
  \begin{center}
  \mbox{\epsfig{figure=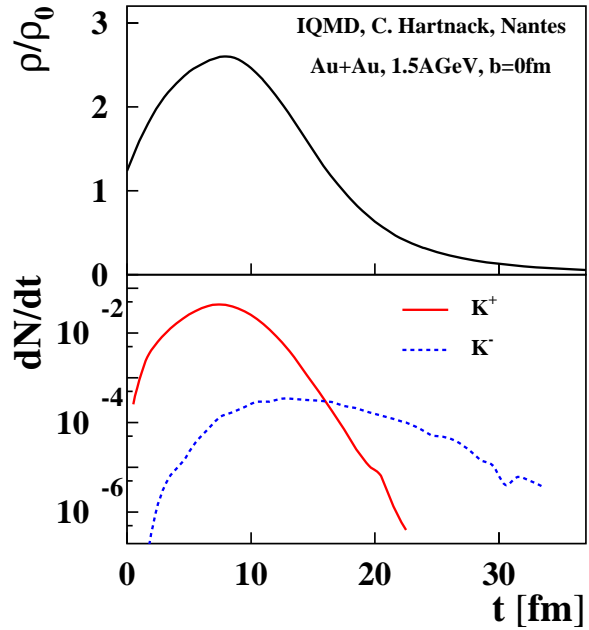,width=9cm,clip=}}
  \caption{The results of an IQMD transport model 
           calculation on the time evolution of a central 
           \mbox{Au+Au} collision \cite{har03_priv}. 
           \underline{Upper panel:} Baryon density 
           $\rho/\rho_{0}$ as a function of the time $t$.
           \underline{Lower panel:} Number of emitted 
           K$^+$ (solid line) and K$^-$ (dashed line)
           per time $dN/dt$ as a function of their 
           production time $t$.} 
  \label{fig_iqmd}
\end{center} 
\end{figure}

The assumption of different emission times for 
K$^+$- and for K$^-$-mesons is supported by a 
transport model calculation using the IQMD code.
Figure \ref{fig_iqmd} shows the results 
of a simulation of central \mbox{Au+Au} collsions 
($b=0$~fm) at $1.5$~AGeV \cite{har03_priv}.  
In the upper panel the time dependence
of the density in the reaction zone is given.
The lower panel shows the rate $dN/dt$ of 
emitted K$^+$ (solid line) and of emitted K$^-$ (dashed line) 
as a function of their production time $t$.
Within this calculation the K$^-$ are 
emitted at a later stage of the reaction 
during the expansion of the reaction volume
while the K$^+$ are emitted during the high-density phase
of the collision.

In summary, we have presented differential cross sections and
phase-space distributions of  K$^+$- and of K$^-$-mesons  produced in
heavy-ion collisions at $1.5$~AGeV. We observed the following
features: (i) The K$^-$/K$^+$ ratio is quite independent of
$\langle A_{part} \rangle$ both for \mbox{Ni+Ni} and 
for \mbox{Au+Au} collisions,  (ii) in
near-central collisions K$^-$-mesons are emitted almost
isotropically whereas K$^+$-mesons exhibit a forward-backward
enhanced emission pattern, and (iii) the inverse slope parameters
are significantly smaller for K$^-$- than for K$^+$-mesons even for
the most central \mbox{Au+Au} collisions. These findings indicate that
(i) the production mechanisms of K$^+$- and of K$^-$-mesons are
correlated by strangeness-exchange reactions, (ii)  K$^-$-mesons
undergo many collisions before leaving the reaction volume and, as a
consequence, (iii) K$^-$-mesons freeze out later than 
K$^+$-mesons.

\end{document}